\documentclass[11pt]{article}
\usepackage{jcappub}
\usepackage{bm}
\usepackage{amsmath,amsthm,amssymb,mathtools}
\usepackage{verbatim}
\usepackage{color}
\usepackage{tabularx}

\def\ref@jnl#1{{\jnl@style#1}}

 \usepackage{url}

\DeclarePairedDelimiter{\ceil}{\lceil}{\rceil}
\DeclarePairedDelimiter{\floor}{\lfloor}{\rfloor}

\renewcommand{\d}{\textrm{d}}

\def\ei{\end{itemize}}
\def\be{\begin{equation}}
\def\ee{\end{equation}}
\newcommand{\bea}{\begin{eqnarray}}
\newcommand{\eea}{\end{eqnarray}}

\newcolumntype{Y}{>{\centering\arraybackslash}X}

\begin{document}

\title{Physics Beyond The Standard Model with Circular Polarization in the CMB and CMB-21cm Cross-Correlation}

\author[a]{Stephon Alexander,}
\author[a]{Evan McDonough,}
\author[b, c]{Anthony Pullen,}
\author[d]{Bradley Shapiro}

 \affiliation[a]{Brown Theoretical Physics Center and Department of Physics, Brown University, Providence, RI, USA}
\affiliation[b]{Department of Physics, New York University University, New York, NY, USA}
\affiliation[c]{Center for Computational Astrophysics, Flatiron Institute, New York, NY, USA}
\affiliation[d]{Department of Physics, Brown University, Providence, RI, USA}

\abstract{Circular polarization is a relatively unexplored realm of CMB physics. Given the substantial community effort towards building next generation CMB polarization experiments, including those which will be sensitive to circular polarization, it behooves theorists to understand the possible sources and relevant physics of circular polarization, as encoded in the Stokes V parameter. In this work we develop and derive the requisite formalism, namely the Boltzmann hierarchy for V-mode scalar, vector, and tensor, anisotropies. We derive the V-mode anisotropies induced by a general source term, and demonstrate how existing proposals for the generation of $V$ can be incorporated as source terms in the Boltzmann hierarchy. A subset of these effects may be correlated with 21cm intensity; we provide a worked example and derive an estimator to extract this information from observations. We conclude by computing the CMB $TV$ cross-correlation generated by axions, and find a relation between $TV$ and $VV$ spectra in axion models. }

\maketitle

\section{Introduction}

The polarization of the cosmic microwave background (CMB) has for decades been a subject of extensive study, and is poised to do so for decades to come. Much of this attention is focused on B-mode polarization, e.g.~as a measure of the energy scale of inflation \cite{Lyth:1996im} or other early universe phenomena \cite{Brandenberger:2011eq}. However, much like the LHC and the energy scale of supersymmetry, there is no strong theory prior on the energy scale of inflation, and hence, nature makes no promise of a detection.  Meanwhile, the CMB circular ``V-mode'' polarization has received very little attention, and is often implicitly taken to have a theory prior of $0$ (see e.g.~\cite{Harrington:2018qpe}).  Similar to B-modes, on the observational front there exist only upper bounds: the V-mode angular power spectrum $\ell (\ell +1) C_{\ell} /(2\pi)$ was constrained by SPIDER to be $\lesssim 10^2 \, \mu\textrm{K}^2$ in the range $33 < \ell < 307$ \cite{Nagy:2017csq}, and by MIPOL to be $\lesssim 10^5 \,\mu\textrm{K}^2$ on larger angular scales \cite{Mainini:2013mja}. Most recently, the CLASS experiment found upper bounds of $0.4 \, \mu {\rm K}^2$ to $13.5\, \mu {\rm K}^2$ in the range $1 \leq \ell \leq 120$ \cite{Padilla:2019dhz}.

From a particle physics perspective, there is a strong science case for studying Stokes V: the requisite imbalance of left- and right-handed photons is indicative of parity violation. We will study one such example of this in detail, namely the interaction of photons with an axion particle. Other works have proposed the generation of $V$ through a variety of mechanisms: Faraday conversion \cite{Cooray:2002nm,Montero-Camacho:2018vgs,Kamionkowski:2018syl,Inomata:2018vbu}, interactions with axions \cite{Agarwal:2008ac,Finelli:2008jv,Alexander:2018iwy,Alexander:2017bxe}, a vector coupling to the photon \cite{Alexander:2008fp} and more general Lorentz violating interactions \cite{Bavarsad:2009hm}, non-commutative geometry \cite{Bavarsad:2009hm,Batebi:2016ocb}, photon-photon scattering via loop corrections in quantum electrodynamics or second-order cosmological perturbation theory \cite{Motie:2011az,Sawyer:2014maa, Sadegh:2017rnr,Montero-Camacho:2018vgs,Kamionkowski:2018syl}, interactions with neutrinos \cite{Mohammadi:2014}, sterile neutrino dark matter \cite{Haghighat:2019rht}, and other fermions \cite{Bartolo:2019}, and polarized Compton scattering due to a non-vanishing bulk velocity of cosmic electrons \cite{Vahedi:2018abn}.
 Interestingly, in contrast to B-modes, the particle physics interactions probed by $V$ are not those at play during inflation, since Thomson scattering in the radiation dominated era quickly and efficiently damps out any primordial $V$ \cite{Alexander:2018iwy}. Thus it is necessarily the physics of the post-inflationary universe which is probed by V-modes.

This latter fact suggests that any beyond the standard model physics which generates $V$ may also affect other probes of the high-redshift universe, and in particular, 21cm cosmology \cite{Furlanetto}. The potential to cross-correlate the CMB anisotropies with 21cm anisotropies opens a wealth of new data to explore. Moreover, any V-generating mechanism that relies upon the production of photons necessarily also contributes to the overall radiation intensity, opening the possibility for cross-correlating the CMB V-mode polarization with the 21cm intensity. This correlator is able to single out the generation of $V$ that occurs not via early universe scatterings, but by traversing cosmology distances.

A primary aim of this work is to emphasize that the theory prior on $V$ as identically $0$ is biased towards the standard $\Lambda$CDM cosmology. There are a wealth of mechanisms beyond the standard model which source $V$ in the CMB, and may be correlated with other observables. In the absence of the detection of B-modes, it may be V-modes that lead the way for the future of CMB polarization. 

To this end, in this work we derive from first principles the Boltzmann hierarchy governing the scalar, vector, and tensor, anisotropies of the Stokes V-parameter. We confirm the suppression of primordial $V$, and find an exact line-of-sight solution that applies in the presence of a general source term. The latter is independent of the details of the underlying mechanism, and takes the form of a sum over all multipole moments. We collect proposals and results in the literature, and map them on to the results derived here. We then turn to 21cm cosmology, and the cross-correlation of the aforementioned V-modes with the 21cm intensity. We build an estimator to extract this information from real data. As a final exercise, we study CMB cross-correlation, namely the $TV$ cross-correlation, and consider the specific case of axions.

The structure of this paper is as follows: In section \ref{Definitions} we introduce definitions and conventions relating to polarization. In section \ref{sec:Boltzmann} we derive the Boltzmann hierarchy for circular polarization, and consider the impact of both primordial $V$ and of a general source term in the Boltzmann equation, and in section \ref{sec:SourceTerms}, we collect V-generating mechanisms from the literature, and demonstrate how they can be incorporated into the Boltzmann hierarchy. In section \ref{sec:21cm} we  compute the cross-correlation with the 21cm intensity, and build an estimator.  In Section \ref{sec:axionsVT} we compute the $TV$ cross-correlation in the CMB. We close in section \ref{sec:discussion} with a discussion of future research directions.

\section{Circular Polarization Preliminaries}
\label{Definitions}

We begin by defining the notation and relevant quantities for the study of circular polarization. We consider an FRW spacetime with scale factor $a(t)$.  For an electromagnetic wave $\vec{E}$ propagating in the $\hat{z}$-direction, the Stokes parameters are defined as
\be\begin{tabular}{ccc}
$\displaystyle I= \frac{1}{a^2}\left( |E_x|^2 + |E_y|^2 \right)$, &\hspace{.2cm}& $\displaystyle V =\frac{i}{a^2}\left( E_x ^* E_y - E_y ^* E_x\right)$, \\[10pt]
$\displaystyle Q=  \frac{1}{a^2}\left( |E_x|^2 - |E_y|^2 \right)$, &\hspace{.2cm}& $\displaystyle U= \frac{1}{a^2}\left( E_x^* E_y + E_y^* E_x\right).$ 
 \end{tabular}\ee
These can be written in a rotated coordinates system $\{x_+,x_-\}$, defined by $\sqrt{2} \hat{x}_\pm = \hat{x} \pm i \hat{y}$,
\be\begin{tabular}{ccc}
$\displaystyle I= \frac{1}{a^2}\left( |E_+|^2 + |E_-|^2 \right)$ , &\hspace{.2cm}& $\displaystyle V =\frac{1}{a^2}\left( |E_+|^2 - |E_-|^2 \right)$, \\[10pt]
$\displaystyle Q=  \frac{1}{a^2}\left( E_+^*E_- + E_-^*E_+ \right)$, &\hspace{.2cm}& $\displaystyle U= \frac{i}{a^2}\left( E_+^*E_- - E_-^*E_+ \right)$,
 \end{tabular}
 \ee
 which makes manifest the nature of $V$ as circular polarization.

The above Stokes parameters have units of intensity, while it is conventional to express the CMB fluctuations as a brightness temperature. Anisotropies in $V$ can be converted to a fractional temperature fluctuation  $\Theta_V$ via the rescaling
\be
\label{deltaTV}
\Theta_V \equiv \frac{V_T}{T} = \frac{V}{I} ,
\ee
where $V_T$ is $V$ in units of temperature, i.e. the brightness temperature perturbation of the circular polarization, while $T$ and $I$ are the background CMB temperature and intensity respectively. In this work we use $\Theta_V$, $V_T$ and $V$ interchangeably, and their distinction should be obvious from the context . 

An essential component in the analysis of CMB polarization is the multipole decomposition. The $V$ polarization at conformal time $\eta$, position $\vec{x}$, and propagation direction $\hat{n}$ is expressed in terms of multipole moments $V_{\ell}^{(m)}$, which we define as
\be \label{multipole}
V(\eta,\vec{x},\vec{n})= \int\frac{\d^3k}{(2\pi)^3}\sum_{m=-2}^2\sum_{\ell\geq \mid m \mid}(2\ell+1)V_\ell^{(m)} (k, \eta)G_\ell^m  ,
 \ee
where we define the Fourier transform as $f(x) = \int {\rm d}^3 k f(k) e^{ikx}$, and
\be 
\label{Glm}
G_\ell^m = (-i)^\ell \sqrt{\frac{4\pi}{2\ell+1}}Y_\ell^m(\hat{n})\exp(i\vec{k}\cdot\vec{x}).
\ee
The power spectrum coefficients $C_{\ell}$ are then given by the two-point correlation function, 
\be
\delta_{\ell \ell'} \delta_{m m' }C_{\ell} ^{VV}  = \langle V_{\ell} ^m(x) V_{{\ell'}} ^{m'*}(y) \rangle,
\ee
For the scalar perturbations $m=m'=0$, this simplifies to
\be
\label{eq:CellVVgeneral}
C_{\ell} ^{VV}(\eta_0) =\int \frac{{\rm d}^3 k}{(2 \pi)^3} {\cal P}^{V} _{\ell}(k)   ,
\ee
with
\be
{\cal P}_{\ell} ^V (k) \delta(k-k') = \langle V_{\ell}(k) V_{\ell}^*(k') \rangle  .
\ee

In this work we also consider the formulation of circular polarization in a spherical basis, as recently done in \cite{Kamionkowski:2018syl}, following the formalism developed in \cite{Dai:2012bc,Dai:2012ma}. In this context, fields are decomposed as
\be
\label{eq:spherical-basis}
\phi(x) = \displaystyle \sum _{\ell m} \int \frac{{\rm d}^3 k}{(2 \pi)^3} \phi^{k} _{\ell m} \Psi ^{k} _{\ell m}(x) ,
\ee
where $\Psi^{k} _{\ell m}$ are eigenstates of total angular momentum, i.e. they satisfy the Helmholtz equation
\be
\left( \nabla^2 + k^2 \right) \Psi_{\ell m}^k = 0 ,
\ee
and are given by
\be
\Psi^{k} _{\ell m} = 4 \pi i^\ell j_{\ell}(kx) Y_\ell^m (\hat{x}) . 
\ee
The power spectrum is related to the two-point correlation function in the usual manner,
\be
\langle \phi^k _{\ell m} \phi ^{k'*} _{\ell' m'} \rangle = \frac{(2 \pi)^3}{k^2} \delta(k - k') \delta_{\ell \ell'} \delta_{m m'} P_{\phi}(k).
\ee
In section \ref{sec:21cm} we will use this formalism to compute the cross-correlation of CMB V-mode polarization with the 21cm temperature anisotropies.

\section{Boltzmann Equations for Circular Polarization}
\label{sec:Boltzmann}

We begin our analysis by establishing the formalism for studying $V$, and the equations of motion that govern the evolution of anisotropies.  We derive the Boltzmann equation for circular polarization brightness temperature perturbations, following closely the analogous calculation for temperature and linear polarization anisotropies \cite{Hu:1997hp}. 

The general form of the Boltzmann equation is given by
\be
\label{eq:Boltz}
 \frac{\rm d }{d \eta} \vec{T} = \vec{C}[\vec{T}] + \vec{G}[h_{\mu \nu}] ,
\ee
where the terms on the right-hand side correspond to collision and gravitational terms respectively, and
\be
\label{vecT} \vec{T} \equiv (T,Q+iU,Q-iU,V),
 \ee 
is the temperature vector. This is a simple generalization of the expression in \cite{Hu:1997hp} to include the $V$ Stokes parameter.

The left-hand side of the Boltzmann equation can be expanded as
\be \frac{\rm d }{d \eta} \vec{T} = \frac{\partial }{\partial \eta} \vec{T} + n^i \nabla_i \vec{T}, \ee
where the gradient term describes the effect of photon free-streaming, given in Fourier space by 
\be
\label{eq:niki}
i n^i k_i= i \sqrt{\frac{4 \pi}{3}} k Y_{1} ^0 .
\ee
Meanwhile, on the right-hand side, gravitational redshift does not affect $V$, and hence the $V$ component of $\vec{G}$ vanishes. This follows from the equivalence principle, which implies that left- and right-handed photons behave identically under gravitational forces. 

The task remains to compute the collision term. In the standard treatment of CMB anisotropies, the collision term is dominated by the effects of Thomson scattering, and we will see that this is also the case for $V$. To do this we follow the method of \cite{Hu:1997hp}: the collision term in the photon rest frame can be derived from the scattering frame via a rotation and de-rotation by the two Euler angles $\alpha$ and $\gamma$.

The angular dependence in the scattering frame can be compactly expressed as,
\be
\begin{pmatrix} \Theta_\parallel\\ \Theta_\perp\\ U\\ V\end{pmatrix}' =
\begin {pmatrix} \cos^2\beta & 0 & 0 & 0\\
0 & 1 & 0 & 0\\
0 & 0 & \cos\beta & 0\\
0 & 0 & 0 & \cos\beta
\end{pmatrix}\begin{pmatrix} \Theta_\parallel\\ \Theta_\perp\\ U\\ V\end{pmatrix}
\ee
where $\Theta_\parallel$ and $\Theta_\perp$ refer to $\Theta$ parallel and perpendicular to the scattering plane respectively, and $\beta$ is the scattering angle. This can be written in terms of $\vec{T}$ as,
\be
\vec{T}' = \mathbf{S}\vec{T} = \frac{3}{4}\begin{pmatrix}
\cos^2\beta+1 & -\frac{1}{2}\sin^2\beta & -\frac{1}{2}\sin^2\beta & 0\\
-\sin^2\beta &\frac{1}{2}(\cos\beta+1)^2 & \frac{1}{2}(\cos\beta-1)^2 & 0\\
-\sin^2\beta &\frac{1}{2}(\cos\beta-1)^2 & \frac{1}{2}(\cos\beta+1)^2 & 0\\
0 & 0 & 0 & 2\cos\beta
\end{pmatrix}\vec{T},
\ee
where we have used the normalization used in \cite{Hu:1997hp}.

We now transform this to to the photon rest frame. Circular polarization $V$ is invariant under rotations, and hence under a rotation parametrized by an angle $\psi$, the vector $\vec{T}$ transforms as $\mathbf{R}(\psi)\vec{T}=\textrm{diag}(1,e^{2i\psi},e^{-2i\psi},1)\vec{T}$.  The properly rotated scattering matrix is then given by
\bea
\mathbf{R}(\gamma)\mathbf{S}(\beta)&& \mathbf{R}(-\alpha)= \\  \frac{1}{2}\sqrt{\frac{4\pi}{5}}
&&\begin{pmatrix}
Y_2^0(\beta,\alpha)+2\sqrt{5}Y_0^0(\beta,\alpha) & -\sqrt{\frac{3}{2}}Y_2^{-2}(\beta,\alpha) & -\sqrt{\frac{3}{2}}Y_2^2(\beta,\alpha) & 0\\
-\sqrt{6}\prescript{}{2}{Y}_2^0(\beta,\alpha)e^{-2i\gamma} & 3\prescript{}{2}{Y}_2^{-2}(\beta,\alpha)e^{-2i\gamma} & 3\prescript{}{2}{Y}_2^2(\beta,\alpha)e^{-2i\gamma} & 0\\
-\sqrt{6}\prescript{}{-2}{Y}_2^0(\beta,\alpha)e^{2i\gamma} & 3\prescript{}{-2}{Y}_2^{-2}(\beta,\alpha)e^{2i\gamma} & 3\prescript{}{-2}{Y}_2^2(\beta,\alpha)e^{2i\gamma} & 0\\
0 & 0 & 0 & \sqrt{15}Y_1^0(\beta,\alpha) 
\end{pmatrix}. \nonumber
\eea
From this one can express the collision term in the photon rest frame as
\be
\vec{C}[\vec{T}]_{rest} = - \dot{\tau} \vec{T} + \dot{\tau}\int \frac{\d\Omega}{4 \pi} \mathbf{R}(\gamma)\mathbf{S}(\beta)\mathbf{R}(-\alpha) \vec{T}(\Omega') .
\ee
Finally, we transform back to the background frame. Similar to  $Q$ and $U$, the circular polarization is not affected by Doppler shifting from the rest frame to the background frame. Hence,  the $V$-component of $\vec{C}[\vec{T}]$ is given by, 
\be
\label{eq:CVrest}
C_{V} = C^{rest}_{V} =  - \dot{\tau} V + \Gamma_V,
\ee
where $\Gamma_V$ is the $V$-component of,
\be
\vec{\Gamma} \equiv \frac{1}{10}\dot{\tau}\int\d\Omega'\sum_{m=-2}^2\textbf{P}^{(m)}(\Omega,\Omega')\vec{T}(\Omega'),
\ee
with
\be
\textbf{P}^{(m)}=\begin{pmatrix}
{Y_2^m}'Y_2^m & -\sqrt{\frac{3}{2}}\;{\prescript{}{2}{Y}_2^m}'Y_2^m & -\sqrt{\frac{3}{2}}\;{\prescript{}{-2}{Y}_2^m}'Y_2^m & 0\\
-\sqrt{6}{Y_2^m}'\prescript{}{2}{Y}_2^m & 3\;{\prescript{}{2}{Y}_2^m}'\prescript{}{2}{Y}_2^m & 3\;{\prescript{}{-2}{Y}_2^m}'\prescript{}{2}{Y}_2^m & 0\\
-\sqrt{6}{Y_2^m}'\prescript{}{-2}{Y}_2^m & 3\;{\prescript{}{2}{Y}_2^m}'\prescript{}{-2}{Y}_2^m & 3\;{\prescript{}{-2}{Y}_2^m}'\prescript{}{-2}{Y}_2^m & 0\\
0 & 0 & 0 & 5{Y_1^m}'Y_1^m
\end{pmatrix},
\ee
and ${Y_\ell^m}'\equiv Y_\ell^m(\Omega')$. From this, one can see that $\Gamma_V$ is given by,
\be
\Gamma_V  = \frac{\dot{\tau}}{2}\sum_{m=-2}^2Y_1^m\int\d\Omega'Y_1^{m*}(\Omega')V(\Omega') ,
\ee 
where $\Omega$ and $\Omega'$ represent directions $\hat{n}$. We now perform a multipole decomposition of $V$ as defined in \eqref{multipole}:
\be
\Gamma_V = \frac{\dot{\tau}}{2}\sum_{m=-2}^2Y_1^m(\Omega)\int\frac{\d^3k}{(2\pi)^3}\sum_{m'=-2}^2\sum_{\ell\geq \mid m' \mid}(2\ell+1)V_\ell^{(m')}\int\d\Omega'Y_1^{m*}(\Omega')G_\ell^{m'}(\vec{x},\Omega',\vec{k}).
\ee
Using the fact that $\int\d\Omega Y_\ell^{m*}Y_{\ell'}^{m'}=\delta_{\ell\ell'}\delta_{mm'}$, this can be further simplified to
\be
\label{eq:GammaV}
\Gamma_V = \int\frac{\d^3k}{(2\pi)^3}\sum_{m=-2}^2\sum_{\ell\geq \mid m \mid}\frac{\dot{\tau}}{2}\delta_{\ell1}(2\ell+1)V_\ell^{(m)}G_\ell^m(\vec{x},\Omega,\vec{k}) .
\ee
This determines the collision term of the Boltzmann equation via equation \eqref{eq:CVrest}.

We can read off the Boltzmann equations of scalar, vector and tensor brightness perturbations, corresponding to $m=0,1,2$ respectively in the multipole decomposition, by combining \eqref{eq:GammaV} with equations \eqref{eq:Boltz}, \eqref{eq:niki}, and \eqref{eq:CVrest},  The respective Boltzmann equations are given by,
\bea \label{eq:VSVT}
{V^{S}} '  + k \mu V^S + \dot{\tau} V^S && = \frac{3}{2}\dot{\tau} V_1 ^S ,\\
{V^{V}}' + k \mu V^V + \dot{\tau} V^V  &&=\frac{3}{2}\dot{\tau} V_1 ^V , \\
{V^{T}}'  + k \mu V^T  + \dot{\tau} V^T &&= 0 ,
\eea
where $V_{1}$ denotes the $\ell=1$ multipole expansion, $\mu \equiv \cos \theta $, and $\{ S,V,T\}$ superscripts denote scalar, vector, and tensor. The scalar and tensor expressions match that given in \cite{Kosowsky:1994cy}, while the Boltzmann equation for vector perturbations has not previously appeared in the literature.  

These equations can be succinctly described by a set of source terms, 
\be
S_V ^{S} = \frac{3}{2}\dot{\tau}   V ^{S}_{1} \;\; ,\;\;  S_V ^V = \frac{3}{2}\dot{\tau}   V ^{V}_{1} \;\; ,\;\; S_V ^T = 0.
\ee
which correspond to the source terms for scalar, vector, and tensor modes of $V$ respectively. These can be compared with the source terms for temperature anisotropies \cite{Hu:1997hp},
\bea
S_{T}^S && = \dot{\tau}\left( \Theta_0^{S}  + v_{B} + \frac{1}{10}\Theta_2 ^S - \frac{\sqrt{6}}{10} E_{2} ^S\right)+\left( k \Psi - \dot{\Phi} \right), \\
S_{T}^V && = \dot{\tau}\left(  v_{B} + \frac{1}{10}\Theta_2 ^V - \frac{\sqrt{6}}{10} E_{2} ^V\right)+\dot{v}, \\
S_{T}^T && = \dot{\tau}\left(  \frac{1}{10}\Theta_2 ^T - \frac{\sqrt{6}}{10} E_{2} ^T\right) - \dot{H}.
\eea
Crucially, the $\Theta_0$ term in $S_{T}^S$ has no counterpart in the Boltzmann equation for $V$. As we will see, this ultimately leads to an uncancelled $\dot{\tau} V_0$ in the equation of motion for $V_0$, resulting in an exponential damping  $V_0 \sim e^{-\tau}$.

\subsection{Multipole Expansion and Boltzmann Hierarchy}
\label{multipoleexpansion}
We now expand the Boltzmann equation to find the Boltzmann hierarchy for the multipoles.  As derived in the previous subsection, the Boltzmann equation for $V$ is given by
\be\label{boltzmanneq}
 \frac{\partial}{\partial\eta}V+ik\sqrt{\frac{4\pi}{3}}Y_1^0V= - \dot{\tau}V + \Gamma_V . 
 \ee
 Using the Clebsch-Gordan relation (see \cite{Hu:1997hp}),
\be
 \sqrt{\frac{4\pi}{3}}Y_1^0Y_\ell^m = \sqrt{\frac{\ell^2-m^2}{(2\ell+1)(2\ell-1)}}Y_{\ell-1}^m+\sqrt{\frac{(\ell+1)^2-m^2}{(2\ell+1)(2\ell+3)}}Y_{\ell+1}^m,
 \ee
the second term on the left-hand side is given by,
\be 
ik\sqrt{\frac{4\pi}{3}}Y_1^0V = \int\frac{\d^3k}{(2\pi)^3}\sum_{m=-2}^2\sum_{\ell\geq \mid m \mid} k\left(\sqrt{(\ell+1)^2-m^2}V_{\ell+1}^{(m)}-\sqrt{\ell^2-m^2}V_{\ell-1}^{(m)}\right)G_\ell^m .
\ee
The Boltzmann equation then becomes,
\begin{align}
0 &= \int\frac{\d^3k}{(2\pi)^3}\sum_{m=-2}^2\sum_{\ell\geq \mid m \mid} \left((2\ell+1)\left({V_\ell^{(m)}}'+\dot{\tau}\left(1-\frac{1}{2}\delta_{\ell1}\right)V_\ell^{(m)}\right)\right. \\
&\hspace{4cm}\left.+k\left(\sqrt{(\ell+1)^2-m^2}V_{\ell+1}^{(m)}-\sqrt{\ell^2-m^2}V_{\ell-1}^{(m)}\right)\right)G_\ell^m. \nonumber
\end{align}
By the linear independence of $G_\ell^m$, we get the Boltzmann hierarchy, i.e. for any $m$ and $\ell$,
\be
 \label{boltzmannhierarchy} 
{V_\ell^{(m)}}'+\dot{\tau}\left(1-\frac{1}{2}\delta_{\ell1}\right)V_\ell^{(m)} +k\left(\frac{\sqrt{(\ell+1)^2-m^2}}{2\ell+1}V_{\ell+1}^{(m)}-\frac{\sqrt{\ell^2-m^2}}{2\ell+1}V_{\ell-1}^{(m)}\right)=0.
 \ee
 This is the V-mode Boltzmann hierarchy; a series of coupled differential equations that directly gives the scalar, vector, and tensor multipole moments ($m=0,1,2$ respectively).   This expression has not been derived in the literature previously.

\subsection{The Fate of Primordial $V$}
\label{analytical}

With the Boltzmann hierarchy in hand, we can immediately study the evolution of $V$ that is present at the beginning of standard cosmology, having been produced in a previous epoch e.g. cosmological inflation, namely \emph{primordial} circular polarization. As emphasized in \cite{Alexander:2018iwy}, any primordial $V$ is exponentially suppressed in the CMB. This can be understood as a straightforward consequence of the suppression of circular polarization by Thomson scattering.  We can see this in detail as follows.

As is the case for temperature \cite{Seljak:1996is}, the scalar modes admit an exact integral solution,
\be \label{standardsolution} V_\ell (\eta) = \int_0^\eta \frac{3}{2}\dot{\tau}e^{-\tau(\eta',\eta)}V_1(\eta')j'_\ell(k(\eta-\eta'))\d\eta', \ee
where
\be \tau(\eta_1,\eta_2) = \int_{\eta_1}^{\eta_2}\dot{\tau}\d\eta , \ee
is the optical depth from $\eta_1$ to $\eta_2$.  This further simplifies in the sudden decoupling approximation, $\dot{\tau}e^{-\tau} =\delta(\eta' - \eta_{LS})$, where the subscript $LSS$ refers to quantity evaluated at last scattering.  In this case, \eqref{standardsolution} becomes,
\be 
\label{Vellsd}
V_\ell \simeq \frac{3}{2}V_1(\eta_{LS})j'_\ell(k(\eta-\eta_{\rm LS })).
\ee
and thus depends very sensitively on the value of $V_{1}$ at last scattering. 

From this, one can compute the spectrum of $V$-mode anisotropies $C_{\ell} ^{VV}$ via equation \eqref{eq:CellVVgeneral}, with $V_1$ determined by the first moments of the Boltzmann hierarchy, 
 \bea
 \label{ell01}
&& V_0 ' + \dot{\tau}V_0 = - k  V_1 \\
&& V_1 ' + \frac{1}{2}\dot{\tau}V_1 = -\frac{2}{3} k  V_2 + \frac{1}{3} k V_0 
\eea
and a similar equation for $V_2$. This system does not admit undamped oscillatory solutions, but instead both $V_0$ and $V_1$ inherit an exponential suppression from the friction term $\dot{\tau}V_{0,1}$.

This behavior can be easily seen by considering long-wavelength modes. Working in a series expansion in $k/\dot{\tau} \ll 1$, and using the fact that $\dot{\tau}$ is decreasing as a function of time, an approximate solution can be found as,
\be
V_1 (\eta) \simeq V_1 (0) e^{- \frac{1}{2}\tau_0 } +\frac{k \eta }{3} e^{- \tau_0} V_0 (0),
\ee
where $\tau_0$ is defined as the optical depth at the beginning of the radiation dominated era, and $V_1(0)$ and $V_0(0)$ are the initial conditions on $V_0$ and $V_1$. The large value of the optical depth $\tau_0$ induces an extreme suppression of $V_1$, and hence by \eqref{Vellsd} the whole tower of $V_{\ell}$. Quantitatively, this suppression is at least of order $10^{10^{20}}$ for standard $\Lambda$CDM cosmology \cite{Alexander:2018iwy}.

This indicates that any observed $V$ must be due to new source terms, and not due to primordial production (e.g.~during inflation). With this in mind, we now study the impact of a completely general source term on $V$-mode polarization, for the moment remaining agnostic as to the physical origins of this effect.

\subsection{Generating V-mode Polarization from a General Source}
\label{sec:CMB}

We now generalize our discussion in section \ref{analytical} to account for general source terms for $V$. In section \ref{sec:SourceTerms} we will see that there are a number of physical processes, in addition to Thomson scattering, which may introduce source terms to the right-hand side of \eqref{boltzmanneq}. Let $S$ be the sum of all these source terms. We may perform a multipole decomposition on $S$, just as we did on $V$, so that the Boltzmann hierarchy becomes
\be 
\label{hierarchywithsource} {V_\ell}'+\dot{\tau}\left(1-\frac{1}{2}\delta_{\ell1}\right)V_\ell +k\left(\frac{\ell+1}{2\ell+1}V_{\ell+1}-\frac{\ell}{2\ell+1}V_{\ell-1}\right) = S_{\ell}.
 \ee
This is a generalization of \eqref{boltzmannhierarchy}; whereas \eqref{boltzmannhierarchy} assumes that there is only an $\ell=1$ source (Thomson scattering), \eqref{hierarchywithsource} permits sources in any mode. 

As with the previous subsection, the hierarchy \eqref{hierarchywithsource} possesses an exact integral solution. However, while \eqref{standardsolution} has only one term corresponding to the $\ell=1$ source term in \eqref{boltzmannhierarchy}, the integral solution for \eqref{hierarchywithsource} will have a sum over \emph{all} $\ell$. It is given by
 \be
  \label{solutionwithsource} 
  V_\ell = \int_0^\eta \left(\frac{3}{2}\dot{\tau}V_1j_\ell'(k(\eta-\eta')) + \sum_{\ell'=0}^\infty S_{\ell'}\sum_{n=0}^{\floor{\frac{\ell'}{2}}} A_n^{\ell'}j_\ell^{(\ell'-2n)}(k(\eta-\eta'))\right)e^{-\tau(\eta',\eta)}\d \eta',
  \ee
 where $A_n^{\ell'}$ is defined recursively as
\be \label{Anldef}
\begin{tabular}{l}
$\displaystyle A_0^{\ell'}=\frac{1}{\Sigma_{00}^{\ell'}}=\frac{(2\ell'+1)!}{2^{\ell'}\ell'!^2}$\\
$\displaystyle A_n^{\ell'}=\frac{-1}{\Sigma_{nn}^{\ell'}}\sum_{p=0}^{n-1}A_p^{\ell'}\Sigma_{np}^{\ell'}$ for $n>0$\end{tabular}
\ee
and we have defined
\be \label{sigmaknldef}
\Sigma^{\ell'} _{kn} = \frac{2^{\ell'-2k}(-1)^{k-n}(\ell'-k-n)!(\ell'-2n)!}{k!(2\ell'-2k-2n+1)!}.
\ee
This provides an exact solution in the presence of any source term.

To demonstrate that \eqref{solutionwithsource} indeed solves \eqref{hierarchywithsource}, one may rely upon the result
\be 
\label{Anelljell}
\sum_{n=0}^{\floor{\frac{\ell'}{2}}} A_n^{\ell'}j_\ell^{(\ell'-2n)}(0) = \delta_{\ell\ell'}, 
\ee
where $A_n^{\ell'}$ and $\Sigma^{\ell'} _{kn}$ are defined in \eqref{Anldef} and \eqref{sigmaknldef}. We present a proof of this  in Appendix  \ref{app:boltzmann}.  With this established, it is easy to check that \eqref{solutionwithsource} solves \eqref{hierarchywithsource}. Applying Leibniz' rule to \eqref{solutionwithsource}, and applying \eqref{Anelljell}, yields
\bea
\displaystyle V'_\ell &&= \frac{1}{2}\dot{\tau}V_\ell\delta_{\ell1} + S_\ell - \dot{\tau}V_\ell \\
&& + k\int_0^\eta \left(\frac{1}{2}\dot{\tau}V_1j_\ell''(k(\eta-\eta')) + \sum_{\ell'=0}^\infty S_{\ell'}\sum_{n=0}^{\floor{\frac{\ell'}{2}}} A_n^{\ell'}j_\ell^{(\ell'-2n+1)}(k(\eta-\eta'))\right)e^{-\tau(\eta',\eta)}\d \eta'. \nonumber
 \eea
The first three terms (i.e., those outside of the integral) trivially cancel with other terms in \eqref{hierarchywithsource}. The last term, as a consequence of the relation,
\be
\label{sphericalbesselimportant} (2\ell+1)j'_\ell(x) = \ell j_{\ell-1}(x)-(\ell+1)j_{\ell+1}(x),
 \ee
 cancels with the $V_{\ell+1}$ and $V_{\ell-1}$ terms in \eqref{hierarchywithsource}. Thus \eqref{solutionwithsource} indeed solves \eqref{hierarchywithsource}

Before we carry on, we should consider the information contained within \eqref{solutionwithsource}. Interestingly, $V_\ell$ depends not only on the corresponding multipole of the source term, but on all multipole moments of the source.  This implies that $V_\ell$ can in principle behave very different than its sources. For example,  in many cases $S_\ell$ will depend on $U_\ell$ or $Q_\ell$, which have well defined trends with respect to $\ell$. However, since $V_\ell$ is a sum over all $S_{\ell'}$, these trends will not necessarily carry over to $V_\ell$.

\section{New Physics and Sources of Circular Polarization}
\label{sec:SourceTerms}

As discussed in section \ref{sec:CMB}, if we wish to model physical processes aside from Thomson scattering, then these will be represented as additional source terms on the right-hand side of \eqref{boltzmannhierarchy}. In this section we collect V-generation proposals in the literature and translate them to explicit forms of the induced source term in the Boltzmann equation. These results are summarized in Table \ref{tab:sources}.

\begin{table*}[h]
\begin{tabularx}{1\textwidth}{|>{\setlength\hsize{1.2cm}\centering}Y|Y|Y|Y|}
   \hline\hline
\bf Section & \bf Physical process & \bf Type of effect & \bf Form of the source term   \\ \hline
\ref{faraday}&  Faraday conversion & Collision effect & proportional to $U_\ell^{(m)}$\\ \hline
\ref{axions} & Axions & Propagation effect & one term for $\ell=m=0$ mode, one term proportional to $V_\ell^{(m)}$, rescaling of all other sources 	\\ \hline
\ref{vector} & Vector coupling to QED & Collision effect & 	linear combination of $U_\ell^{(m)}$ and $V_\ell^{(m)}$ 	\\ \hline
\ref{qed} & One-loop QED (photon-photon scattering) 	& Collision effect  & proportional to $U_\ell^{(m)}$	\\ \hline
\ref{photonfermion} & Photon-fermion scattering 	& Collision effect & 	mixing of $V_\ell^{(m)}$ with different multipoles of $U$ and $Q$, including vector and tensor modes  	\\ \hline
\ref{spacetime} & Non-commutative spacetime	 & Propagation effect & 	linear combination of $U_\ell^{(m)}$ and $V_\ell^{(m)}$ 	\\ \hline
   \hline\hline
\end{tabularx}
\caption{Sources of Circular Polarization}
\label{tab:sources}
\end{table*}

\subsection{Faraday Conversion}
\label{faraday}

Early work on circular polarization \cite{Cooray:2002nm} introduced Faraday conversion as an important tool for understanding varied sources. Recalling,
\be
\begin{tabular}{c}
$V=\frac{2}{a^2}\rvert E_x\rvert\rvert E_y \rvert\sin\Delta\phi$ ,\\
$U=\frac{2}{a^2}\rvert E_x\rvert\rvert E_y \rvert\cos\Delta\phi$ ,
\end{tabular}
\ee
we see that if the amplitudes $\rvert E_x\rvert$ and $\rvert E_y \rvert$ are constant, then it follows that $V$ and $U$ are related via
\be
 V'=U\frac{\d\Delta\phi}{\d\eta}=2U\frac{\d\Delta\phi_{FC}}{\d\eta} ,
 \ee
where, following the convention of \cite{Cooray:2002nm}, $\Delta\phi_{FC}=\frac{1}{2}\Delta\phi$. This implies that any process that introduces a phase difference between the $x$ and $y$ components will cause $V$ to change at a rate proportional to $U$. This phenomenon is known as Faraday conversion\footnote{Note that Faraday conversion should not be confused with Faraday \emph{rotation}, also discussed in \cite{Cooray:2002nm}.}. It can be incorporated in the Boltzmann hierarchy via the source term,
\be 
\label{Sphic}
S_{\ell}^{(m)} =  2U_\ell^{(m)}\frac{\d\Delta\phi_{FC}}{\d\eta}. 
\ee
We will discuss several sources of Faraday conversion below, regardless of sourcing phenomenon.

One source of Faraday conversion is that caused by a magnetized relativistic plasma \cite{Cooray:2002nm}, in which
\be 
\Delta\phi=\frac{e^4\lambda^3}{\pi m_e^3c^5}\frac{\beta-1}{\beta-2}\int\d l \, n_r \gamma_{\textrm{min}}\rvert\bold{B}\rvert^2(1-\mu^2).
\ee
where $\bold{B}$ is the magnetic field, $m_e$ is the electron mass, $\mu$ is the cosine of the angle between $\bold{B}$ and the line of sight, $n_r$ is the number density of relativistic particles, $\beta$ is an index for the distribution of the particles in terms of their Lorentz-factor $\gamma$ \cite{Cooray:2002nm},  $\gamma_{min}$ is the minimum value of $\gamma$, and $\lambda$ is the wavelength of radiation. The total impact of this source on $V$ (in units of $T_{CMB}$) is estimated to be on the order of $10^{-9}$ at 10 GHz.

As an additional example of Faraday conversion, \cite{Montero-Camacho:2018vgs} showed that if hydrogen atoms become spin-polarized, they may induce a birefringence between the $x$ and $y$ directions. This spin-polarization is caused by Balmer radiation during the epoch of recombination, and by 21cm radiation during the cosmic dawn era (in both cases ultimately coming from other nearby hydrogen atoms). Specifically, this spin-polarization induces a phase change at a rate of
\be
 \frac{\d\Delta\phi}{\d\ln a}=-C\left(1+z\right)^{\frac{1}{2}}\left(\mathcal{P}_{2,2}+\mathcal{P}_{2,-2}\right) \frac{100\textrm{ GHz}}{\nu_{\textrm{today}}},
 \ee
where $C=3.13\times10^{-3}$ during the cosmic dawn and $C=2.45\times10^{-3}$ during recombination, and the alignment tensor $\mathcal{P}_{2,m}$ is given in \cite{Montero-Camacho:2018vgs}. The total impact of this source on $V$ is estimated to be (in units of $T_{CMB}$) on the order of $10^{-16}$ during the cosmic dawn and $10^{-19}$ during recombination.

\subsection{Axions}
\label{axions}

Axions naturally interact with photons according to the Lagrangian
\be
\label{eq:Lintaxions}
\mathcal{L}_{int}= -\frac{g_\phi}{4}\phi F_{\mu \nu} \tilde{F}^{\mu \nu}
\ee
where $F^{\mu \nu} = \epsilon^{\mu \nu \sigma \rho}F_{\sigma \rho}$ is the Hodge dual of Maxwell field strength tensor $F$, $\phi$ is the axion field, and $g_{\phi}$ is the coupling constant. This induces two qualitatively distinct effects: axion-photon scattering, discussed in section \ref{vector}, and a modified dispersion relation for photons \cite{Finelli:2008jv}:
\be \label{dispersion} A''_\pm + (k^2\pm g_\phi\phi'k)A_\pm = 0 \ee
where $A_\pm$ are (Fourier-transformed) components of the electromagnetic four-potential. The resulting $A_{\pm}$ depend sensitively on time-dependence $\phi$, which is model-dependent.

For the moment, we work in full generality. Let us write the solution as
\be
 \label{gammadef} A_\pm = A_{\pm0}e^{ik\int\gamma_\pm\d\eta}. 
 \ee
Note that any solution to \eqref{dispersion} can be written in this way; this follows from the fact that if $f_\pm(\eta)$ solves \eqref{dispersion}, then
\be \gamma_\pm=\frac{1}{ik}\frac{\d}{\d\eta}\ln f_\pm(\eta). \ee
In the trivial case of $\phi'=0$, we get $\gamma_\pm=1$; this is the case where there is no photon-axion interaction. If instead $\phi'$ is constant, e.g. during a slow-roll phase, then 
\be
\label{gpmaxion}
\gamma_\pm = \sqrt{1\pm \frac{g_\phi\phi'}{k}}.
\ee  
Taking the derivative of \eqref{gammadef}, we get that
\be
 \rvert A'_\pm\rvert = \rvert k\gamma_\pm A_{\pm0}\rvert e^{-k\int\operatorname{Im}(\gamma_\pm)\d\eta}.
 \ee
But\ recall that when there is no photon-axion interaction, $\gamma_\pm=1$, and so in that case $\rvert A'_\pm\rvert = \rvert kA_{\pm0}\rvert$. Defining $A_{\pm\textrm{w/o}}$ as the value that $A_{\pm}$ would have if there were no photon-axion interaction, we get
\be 
\rvert A'_\pm\rvert = \rvert\gamma_\pm\rvert e^{-k\int\operatorname{Im}(\gamma_\pm)\d\eta}\left\rvert A'_{\pm\textrm{w/o}}\right\rvert.
\ee
We define $\Gamma_\pm = \rvert\gamma_\pm\rvert^2 e^{-2k\int\operatorname{Im}(\gamma_\pm)\d\eta}$, so that
\be 
\label{scaling} \rvert A'_\pm\rvert^2 = \Gamma_\pm\left\rvert A'_{\pm\textrm{w/o}}\right\rvert^2.
\ee
One can then compute the source term, and applying the definition of the Stokes parameters, one arrives at
\be 
V'=\frac{\Gamma'_++\Gamma'_-}{\Gamma_++\Gamma_-}V+\frac{\Gamma_-\Gamma'_+-\Gamma_+\Gamma'_-}{\Gamma_++\Gamma_-}T+\frac{\Gamma_+-\Gamma_-}{2}T'+\frac{\Gamma_++\Gamma_-}{2}V'_{\textrm{w/o}},
 \ee
where $T$ is the temperature fluctuation, and $V_{\rm w/o}$ is the $V$ generated in the absence of axions. 

Thus we see that axions introduce four changes to $V'$: it introduces one term proportional to $V$ and another term proportional to $T$, it causes all sources of $T$ to become sources of $V$ (scaled by $(\Gamma_+-\Gamma_-)/2$), and it causes all of $V$'s pre-existing (i.e., non-axion) sources to be scaled by $(\Gamma_++\Gamma_-)/2$. Performing a multipole decomposition, we find that the right-hand side of the Boltzmann hierarchy \eqref{boltzmannhierarchy} becomes
\bea
\label{axionsources} \displaystyle S_{V\ell}^{(m)} = &&\frac{\Gamma'_++\Gamma'_-}{\Gamma_++\Gamma_-}V_\ell^{(m)}+\frac{\Gamma_-\Gamma'_+-\Gamma_+\Gamma'_-}{\Gamma_++\Gamma_-}T_\ell^{(m)} \\
&&+\frac{\Gamma_+-\Gamma_-}{2}\sum S_{T\ell}^{(m)}+\frac{\Gamma_++\Gamma_-}{2}\sum \hat{\mathcal{S}}_{V\ell}^{(m)}. \nonumber
\eea
where ${\cal S}_{T}$ and $\hat{\cal S}_V$ are the source-term for $T$ and non-axion sources for $V$ respectively, and the sum is over all such source terms. Additionally, because the $\dot{\tau}$ terms in \eqref{boltzmannhierarchy} are the source term for Thomson scattering, they also are rescaled by $\frac{\Gamma_++\Gamma_-}{2}$.

The above results are entirely general, but one specific case is worth considering \cite{Finelli:2008jv}. This is the adiabatic model, in which $\frac{g_\phi\phi''}{k^2}\ll1$ for all modes observable in the CMB, and $\sqrt{1\pm\frac{g_\phi\phi'}{k}}\in\mathbb{R}$. Since $\phi''$ is insignificant in this model, $\gamma_\pm=\sqrt{1\pm\frac{g_\phi\phi'}{k}}$, and so $\Gamma_\pm=1\pm\frac{g_\phi\phi'}{k}$ (see section IV A of \cite{Finelli:2008jv} for more details). Thus \eqref{axionsources} becomes
\be 
\label{adiab}
S_{V\ell}^{(m)}=\frac{g_\phi\phi''}{k}T_\ell^{(m)}+\frac{g_\phi\phi'}{k}\sum S_{T\ell}^{(m)}+\sum \hat{\mathcal{S}}_{V \ell}^{(m)}, 
\ee
where again ${\cal S}_{T}$ and $\hat{\cal S}_V$ are the source-term for $T$ and non-axion sources for $V$ respectively, and the sum is over all such source terms. We see that in this adiabatic model, non-axion sources are not scaled, while all sources of $T$ are introduced as sources of $V$, and there is a direct sourcing of $V$ by $T$ (i.e. the first term in the above).

\subsection{General Vector Coupling to Photons}
\label{vector}
A generalization of the axion-gauge field interaction is to promote $\partial_{\mu}\phi$ to a general vector $T_{\mu}$ \cite{Alexander:2008fp} and consider the Lagrangian,
\be
\mathcal{L}_T=g_T\epsilon^{\mu\nu\alpha\beta}A_\mu T_\nu F_{\alpha\beta} ,
\ee 
where $g_T$ is a coupling constant. This causes polarization modes to rotate in to each other, and the sourced V-modes are given by,
\be V'=\frac{g_T^2}{ak}\left(\chi(\hat{k})U+\zeta(\hat{k})Q\right) ,
\ee
where $\chi$ and $\zeta$ are functions of the stress-energy tensor and the polarization four-vectors. There are similar relations for the rotation of $V$ into $U$ and $Q$, however these may be safely ignored to a good approximation, since the $V$ contribution is vastly subdominant to standard contributions to $U$ and $Q$. To incorporate this effect in the Boltzmann hierarchy, the source term is given by,
\be 
\label{smehierarchygenvec} S_\ell^{(m)} = \frac{g_T^2}{ak}\left(\chi(\hat{k})U_\ell^{(m)}+\zeta(\hat{k})Q_\ell^{(m)}\right).
 \ee
Note that the axion model considered in the previous section is actually a specific case of the model in this section; the two are equivalent  with the identification $T_\nu=\partial_\nu\phi$. In that case, one should incorporate both the effects of this section and of the previous section, including both as source terms in the Boltzmann equation. The latter is a propagation effect, while \eqref{smehierarchygenvec} is a collision effect, and the two can happen simultaneously.

\subsection{One-loop QED (Second Order Cosmological Perturbation Theory)}
\label{qed}

An additional possibility is that circular polarization can be generated at second order in cosmological perturbation theory, via photon-photon scattering. This was studied by \cite{Montero-Camacho:2018vgs}, and expanded on in \cite{Kamionkowski:2018syl}, which found a Faraday conversion
\be
 \frac{\d \Delta \phi_{FC}}{\d\ln a}=8.7\times10^{-8}\left(\frac{\nu_{\textrm{today}}}{100\textrm{ GHz}}\right)\left(\frac{1+z}{1000}\right)^{\frac{7}{2}}\frac{\textrm{Re}\,a_{22}^E}{10^{-6}}.
 \ee
where $a_{2m}^E$ is the local quadropole moment. The corresponding source term in the Boltzmann hierarchy follows from \eqref{Sphic}.  

For the sake of completeness, we note differing results in the literature:  \cite{Motie:2011az} found $V' \propto UQ$; \cite{Sawyer:2014maa}, while \cite{Sadegh:2017rnr} found rotations of both $U$ and $Q$ into $V$, i.e. of the form $V' \propto U+Q$. For a detailed comparison and discussion see \cite{Montero-Camacho:2018vgs}. 



Photon-photon interactions do represent a somewhat different situation than the phenomena we have considered thus far, since they are not new physics \emph{per se}. Whereas we do not know whether axions and non-commutative spacetimes actually exist, we do know for a fact that photons do interact with other photons. Thus, any circular polarization induced by photon-photon interactions will represent a baseline $V$, analogous to the neutrino floor in direct detection searches for dark matter, onto which the effects of other phenomena might be added.

\subsection{Photon-Fermion Interactions}
\label{photonfermion}
In a similar vein to the previous section, \cite{Mohammadi:2014} examines the effect of photon-neutrino scattering\footnote{We note that, following the first footnote in \cite{Montero-Camacho:2018vgs}, it is not clear that photon-neutrino scattering should source circular polarization. \cite{Montero-Camacho:2018vgs} presents an elegant argument that there should be no neutrino-induced $V$, which is clearly at odds with the result of \cite{Mohammadi:2014}. We therefore encourage the reader to maintain some skepticism with regards to \eqref{neutrinoscatter}.}, finding
\be 
\label{neutrinoscatter} V'=a\frac{\sqrt{2}\alpha G_F}{3\pi k^0}\int\d\vec{q}\, n_\nu(\vec{x},\vec{q})\left[\left(\left(\vec{q}\cdot\vec{\epsilon}_1\right)^2-\left(\vec{q}\cdot\vec{\epsilon}_2\right)^2\right)Q-2\left(\vec{q}\cdot\vec{\epsilon}_1\right)\left(\vec{q}\cdot\vec{\epsilon}_2\right)U\right], 
\ee
where $n_{\nu}$ is the number density of neutrinos, \cite{Bartolo:2019} considers the more general case of photon-fermion scattering, and \cite{Haghighat:2019rht} considers sterile neutrino dark matter. These cases lead to a similar structure, where $U$ and $Q$ both source $V$:
\be
 \label{generalscatter}
V' = c_Q Q+ c_U U 
\ee
where the coefficients $c_{Q,U}$, given in  \cite{Haghighat:2019rht}, themselves have an angular dependence.
%

In this case, one cannot deduce the source terms for the multipole moments by simply transforming $Q \to Q_\ell^{(m)}$ and $U \to U_\ell^{(m)}$, as we have done in previous sections, since $c_{Q}$ and $c_{U}$ have their own angular dependence. What we must therefore do is to write these coefficients as linear combinations of spherical harmonics:
\begin{align}
V' &= \left(\sum_{\ell,m}c_{Q\ell,m}Y_\ell^m(\theta,\phi)\right)Q+\left(\sum_{\ell,m}c_{U\ell,m}Y_\ell^m(\theta,\phi)\right)U \nonumber\\
&= \int\frac{\d^3k}{(2\pi)^3}\sum_{\ell,m}\sum_{\ell',m'}(2\ell'+1)\left(c_{Q \ell,m}Q_{\ell'}^{(m')}+c_{U \ell,m}U_{\ell'}^{(m')}\right)Y_\ell^m(\theta,\phi)G_{\ell'}^{m'}(\theta,\phi).
\end{align}
Simplifying this further can be done via manipulation of identities and Clebsch-Gordan coefficients.

However, even without detailed calculations, one can see an interesting consequence of this angular dependence. A term proportional to $Y_L^{m+m'}$ in the expression for $V'$ corresponds to a source term in the $V_L^{(m+m')}$ Boltzmann hierarchy equation. We have seen that the expression for $V'$ will incorporate terms proportional to $Q_{\ell'}^{(m')}Y_L^{m+m'}$ and $U_{\ell'}^{(m')}Y_L^{m+m'}$, so that $V_L^{(m+m')}$ is coupled to $Q_{\ell'}^{(m')}$ and $U_{\ell'}^{(m')}$. Since $\ell'$ and $m'$ need not be equal to $L$ and $m+m'$, what we get is that a $V$ multipole moment can be coupled to $Q$ and $U$ multipole moments with different values of $\ell$ and $m$. A tensor perturbation of $Q$ can affect a scalar perturbation of $V$, for example.


\subsection{Non-Commutative Spacetime}
\label{spacetime}
Non-commutative space and spacetime arise in string theory \cite{Seiberg:1999vs}, and are thought to arise in  quantum gravity more generally. Stated formally, this corresponds to a non-zero commutator,
\be \theta^{\mu\nu}=-i\lbrack\hat{x}^\mu,\hat{x}^\nu\rbrack, \ee
where $\theta^{\mu\nu}$ is a real, constant, antisymmetric matrix.  This non-commutative geometry causes $Q$ and $U$ to convert into $V$ \cite{Batebi:2016ocb}, as
\be
V'= d_Q Q+ d_U U,
  \ee
where $d_{Q/U}$ are functions of angle and the polarization vector. Similar to photon-fermion interactions of the previous subsection, the angular dependence of the coefficients gives a non-trivial structure to the source term. The resulting  $C^{VV}_\ell$ is given by \cite{Batebi:2016ocb} as roughly on the order of $\textrm{nK}^2$ if the energy-scale of non-commutativity $\Lambda$ is $\sim10\textrm{ TeV}$, or on the order of $\mu\textrm{K}^2$ if $\Lambda\sim1\textrm{ TeV}$.



\section{CMB-21cm Cross Correlation}
\label{sec:21cm}

We now shift our focus away from the CMB, and towards a broader perspective on the future of cosmology.

A promising new observational probe of cosmological physics is the 21cm signal; produced by hyperfine splitting in the hydrogen atom, this is a tracer of the intergalactic medium (IGM). For a detailed review see  \cite{Furlanetto}. This probes both the cosmic dark ages and the epoch of reionization, two periods in the universe's history about which we currently have very little information. In addition to circumventing the cosmic variance limitations of CMB probes, measurements of the power spectrum from the dark ages can in principle be more accurate than those from the CMB, because during the dark ages the IGM is not affected by photon diffusion.  At lower redshifts, the 21cm signal probes cosmic structure, as after reionization the 21cm signal traces dark matter halos rather than the IGM. Observations of the global 21cm signal were first reported by the EDGES experiment \cite{Bowman:2018yin}, and many new experiments will be going after both the global signal \cite{2013AAS...22122905L,Bernardi:2018ysg,Ansari:2018ury} and anisotropies, e.g. LOFAR \cite{2013A&A...556A...2V}, MWA \cite{2013PASA...30....7T}, PAPER \cite{2010AJ....139.1468P},  CHIME \cite{Bandura:2014gwa}, HIRAX \cite{Newburgh:2016mwi}, HERA \cite{DeBoer:2016tnn}, and SKA \cite{Bull:2018lat}. See \cite{Trott:2019lap} for a detailed review.

The high energy physics prospects of 21cm cosmology range from precision measurements of primordial non-Gaussianity \cite{Munoz:2015eqa,Sekiguchi:2018kqe,Ansari:2018ury}, or constraining or observing the velocity-dependence of dark matter scattering cross section \cite{Munoz:2018pzp,Munoz:2018jwq,Barkana:2018cct}, to a probe of cosmic strings \cite{Brandenberger:2010hn,McDonough:2011er}. This is in addition to constraining the dark energy equation of state, dark matter-dark energy interactions, and the standard $\Lambda$CDM parameters.

If there is CP-violating physics at play during the dark ages of epoch of reionization, this may generate circular polarization of the 21cm emission.  However, instead of going after the circular polarization of the 21cm signal directly, here we propose to isolate CP-violating effects using the cross-correlation of the \emph{intensity} of the 21cm signal with the Stokes $V$ of the CMB. This is conceivably within reach for next generation cosmology experiments. However, we do note that 21cm circular polarization is in itself an interesting observable, and can probe primordial gravitational waves \cite{Hirata:2017dku,Mishra:2017lpz}.

We emphasize that despite being generated by parity-violating processes, the V Stokes parameter of the CMB is a \emph{scalar} \cite{Kosowsky:1994cy}, and the multipole expansion into $Y_{\ell} ^m$ enjoys the same parity properties as that for $T$. This allows for non-vanishing $TV$ cross-correlation, and similarly, CMB-V-21cm-I. This is in contrast to $B$-mode polarization, which is expanded in spin-2 basis functions and has opposite parity to $T$, implying the TB cross-correlation vanishes in the absence of additional parity violating effects, such as chiral gravitational waves \cite{Lue:1998mq}. Incidentally, many of the effects that lead to $V$ will also source chiral gravitational waves (e.g. axions \cite{Sorbo:2011rz,Ferreira:2014zia,Ferreira:2015omg,Namba:2015gja,Adshead:2013qp, Adshead:2013nka, Maleknejad:2016qjz,Maleknejad:2014wsa, Maleknejad:2016dci,Caldwell:2017chz,Dimastrogiovanni:2016fuu,Bielefeld:2015daa,McDonough:2018xzh}), and thus one additionally expects non-vanishing $TB$, $EB$, and $VB$ in these models.

Among these possibilities, the cross correlation of $V$ and the 21cm signal is uniquely positioned to probe the propagation of photons across cosmological distances. While cross-correlations of $V$ with any signal originating before last scattering, such as $V$ itself, encode information about both collision and propagation effects, a cross-correlation between $V$ and a signal arising \emph{after} last scattering will encode only information about the propagation effects. 

With this in mind, here we develop the CMB-21cm cross-correlation in the model of \cite{Kamionkowski:2018syl}, and compute an estimator for extracting this from data.

\subsection{Faraday Conversion}

\label{21cmtoy}

Here we consider the model of \cite{Kamionkowski:2018syl}. In this scenario, circular polarization is generated by the propagation of light through a medium with an anisotropic index of refraction tensor,
\be
n_{ij} (\vec{x}) \propto \left( \nabla_i \nabla_j - \frac{1}{3} g_{ij} \nabla^2 \right) \hat{\delta}(\vec{x}) ,
\ee
where $\hat{\delta}$ is the matter perturbation; typically written as $\delta$, we have added a hat to differentiate it from the Kronecker delta. This leads to Faraday conversion along the lines discussed in section \ref{faraday}.

The resulting circular polarization is given by,
\be
V(\hat{n}) = \epsilon_{ab} P^{ac} \Phi^b _c, 
\ee
where $P$ is the linear polarization matrix, 
\begin{equation}
     P_{ab}(\hat n) \equiv \frac{1}{\sqrt{2}}\left( \begin{matrix}
     Q(\hat n) & U(\hat n) \\ U(\hat n) &
     -Q(\hat n) \end{matrix} \right) 
\end{equation}
and $\Phi_{ab}$ is a phase shift tensor, 
\be
\Phi_{ab} (\hat{n}) \equiv \frac{1}{\sqrt{2}}\left(\begin{matrix}
     \phi_Q(\hat n) &
     \phi_U(\hat n) \\ \phi_U(\hat n) &
     -\phi_Q (\hat n) \end{matrix} \right). 
\ee
In both expressions $\hat{n}$ is a direction that specifies a point on the celestial sphere. We expand these in multipoles as,
\be
\Phi_{ab} (\hat{n}) = \sum_{\ell m} \Phi_{\ell m} Y ^{\rm E} _{(\ell m)ab}(\hat{n}) ,
\ee
and similarly for $P_{ab}(\hat{n})$, where $Y^E$ are the E-mode tensor spherical harmonics (defined in \cite{Kamionkowski:2016}).

One can straightforwardly compute the spherical harmonic coefficients for $V$, and these are given by
\be
V_{\ell m} = \displaystyle \sum_{\ell_1 \ell_2 m_1 m_2} P_{\ell_1 m_1}\Phi_{\ell_2 m_2} G^{\ell m} _{\ell_1 m_1 \ell_2 m_2},
\ee
where $G$ is defined in terms of Wigner-3j functions  \cite{Kamionkowski:2018syl}.  We note that the $V_{\ell m}$ are different from $V_{\ell}^{(m)}$ used in previous sections: $V_\ell^{(m)}$ refer to a Fourier mode of a multipole moment, as per \eqref{multipole}, while the  $V_{\ell m}$ above are position space multipole moments, defined by the decomposition,
\be
\label{VlmYlm}
 V(\hat{n})=\sum_{\ell m}V_{\ell m}Y_{\ell m}(\hat{n}).
  \ee
 We also note that the definition of $V_\ell^{(m)}$ includes scaling factors via the definition of $G_\ell^m$, which are not present in the definition of $V_{\ell m}$.

For the specific case of birefringence due to spin polarizations of hydrogen atoms, one finds  \cite{Kamionkowski:2018syl} that 
the phase shift is determined by the matter perturbation, as
\be
 \Phi_{\ell m}^k=2\sqrt{15\pi}i^\ell\frac{p}{N_\ell}\int\d\chi W(\chi)\frac{j_\ell(k\chi)}{(k\chi)^2}\hat{\delta}_{\ell m}^k, 
 \ee
where again $\hat{\delta}$ denotes the matter perturbation, $p$ and $W$ are given in \cite{Montero-Camacho:2018vgs}, and  $N_l \equiv \sqrt{2 (l-2)! /(l+2)!}$. It is important to note that $\Phi_{\ell m}^k$ is not a coefficient of the expansion given in \eqref{eq:spherical-basis}; rather, it is the contribution to $\Phi_{\ell m}$ from a given value of $k$, i.e.
\be
 \Phi_{\ell m}=\int\frac{k^2\d k}{(2\pi)^3}\Phi_{\ell m}^k. 
 \ee
Here $\Phi_{\ell m}^k$ is a notational exception; every other quantity with a subscript of $\ell m$ and superscript of $k$ will indicate a coefficient in the spherical basis, as per \eqref{eq:spherical-basis}.

We now consider the 21cm signal, $\tilde{\delta}_{\ell m}^k$. We assume that the 21cm is a biased tracer of the matter distribution, such that
\be \tilde{\delta}_{\ell m}^k = a_k\hat{\delta}_{\ell m}^k,\ee
where $a_k$ is a redshift-dependent constant of proportionality. The cross-correlation with the matter density perturbation is then given by,
\be \langle\hat{\delta}_{\ell m}^k\tilde{\delta}_{\ell'm'}^{k'*}\rangle = a_k^* \langle\hat{\delta}_{\ell m}^k\hat{\delta}_{\ell'm'}^{k'*}\rangle = a_k^* \delta_{\ell \ell'}\delta_{mm'}\delta_{kk'}P_{\hat\delta\hat\delta}(k),\ee
where $P_{\hat\delta\hat\delta}(k)$ is the matter perturbation power spectrum. It follows that the circular polarization-matter perturbation cross correlation is then given by
\begin{align}
\langle \Phi_{\ell m}^k\tilde{\delta}_{\ell'm'}^{k'*} \rangle &= 2\sqrt{15\pi}i^\ell\frac{p}{N_\ell}\int\d\chi W(\chi)\frac{j_\ell(k\chi)}{(k\chi)^2}\langle\hat{\delta}_{\ell m}^k\tilde{\delta}_{\ell'm'}^{k'*}\rangle \nonumber\\
&= 2\sqrt{15\pi}i^\ell\frac{p}{N_\ell}a_{k'}^* \delta_{\ell\ell'}\delta_{mm'}\delta_{kk'}P_{\hat\delta\hat\delta}(k)\int\d\chi W(\chi)\frac{j_\ell(k\chi)}{(k\chi)^2},
\end{align}
and the $\Phi\tilde{\delta}$ angular power spectrum is
\be\label{clphidelta} C_\ell^{\Phi\tilde{\delta}} = \int\frac{k^2\d k}{(2\pi)^3}\frac{k'^2\d k'}{(2\pi)^3}\langle \Phi_{\ell m}^k\tilde{\delta}_{\ell m}^{k'*}\rangle = 2\sqrt{15\pi}i^\ell\frac{p}{N_\ell}\int\frac{k^2\d k}{(2\pi)^3}a_{k}^* P_{\hat\delta\hat\delta}(k)J_\ell^W(k), \ee
where we have defined
\be J_\ell^W(k) = \int\d\chi W(\chi)\frac{j_\ell(k\chi)}{(k\chi)^2}. 
\ee
We note the factor of $i^\ell$ is a reflection of the choice of basis functions for the multipole expansion \eqref{VlmYlm}; the physical observable $C(\theta)$ is real.

We will use \eqref{clphidelta} to compute the $V\tilde{\delta}$ angular power spectrum. The cross correlation is:
\bea
\langle V_{\ell m}{\tilde{\delta}_{\ell'm'}}^{k'*}\rangle &&= \sum_{\ell_1m_1\ell_2m_2}\langle P_{\ell_1m_1}\Phi_{\ell_2m_2}{\tilde{\delta}_{\ell'm'}}^{k'*}\rangle G_{\ell_1m_1\ell_2m_2}^{\ell m} \nonumber\\
&&= \sum_{\ell_1m_1\ell_2m_2}\left(\langle P_{\ell_1m_1}\rangle\langle\Phi_{\ell_2m_2}{\tilde{\delta}_{\ell'm'}}^{k'*}\rangle + \langle\Phi_{\ell_2m_2}\rangle\langle P_{\ell_1m_1}{\tilde{\delta}_{\ell'm'}}^{k'*}\rangle + \langle{\tilde{\delta}_{\ell'm'}}^{k'*}\rangle\langle P_{\ell_1m_1}\Phi_{\ell_2m_2}\rangle\right) G_{\ell_1m_1\ell_2m_2}^{\ell m} \nonumber\\
&&= \sum_{\ell_1m_1\ell_2m_2}\langle P_{\ell_1m_1}\rangle\langle\Phi_{\ell_2m_2}{\tilde{\delta}_{\ell'm'}}^{k'*}\rangle G_{\ell_1m_1\ell_2m_2}^{\ell m} \nonumber\\
&&= \int\frac{k^2\d k}{(2\pi)^3}\sum_{\ell_1m_1}\langle P_{\ell_1m_1}\rangle\langle\Phi_{\ell'm'}^k{\tilde{\delta}_{\ell'm'}}^{k'*}\rangle G_{\ell_1m_1\ell'm'}^{\ell m}.
\eea
In second line we have approximated the three-point function by considering one of $P$, $\Phi$, or $\tilde{\delta}$, to be a long-wavelength fluctuation, such that it is an effective background for the cross-correlations. In the second-to-last line we have used the fact that, absent any effects beyond the Faraday conversion of $P$ into $V$, $P$ and $\tilde{\delta}$ are uncorrelated, as are $P$ and $\Phi$.

When considering the entire sky\footnote{As in the previous sections, here we equate ensemble and spatial averages, by the Ergodic hypothesis}, $\langle P_{\ell_1m_1} \rangle$ vanishes. Thus in the above approximation of three-point function, the all-sky $\langle V_{\ell m}{\tilde{\delta}_{\ell'm'}}^{k'*}\rangle$ also vanishes. However, on any individual patch of the sky, $\langle P_{\ell_1m_1} \rangle$ need not be 0. The average value of $P_{\ell_1m_1}$ on such a patch is given by the rms, $\sqrt{\langle P_{\ell_1m_1}^2\rangle}$, which we will define as $\sigma_{\ell_1m_1}$. Therefore, the angular power spectrum is given by
\begin{align}
C_{\ell m}^{V\tilde{\delta}} &= \int\frac{k'^2\d k'}{(2\pi)^3}\langle V_{\ell m}{\tilde{\delta}_{\ell m}}^{k'*}\rangle \nonumber\\
&=\int\frac{k^2\d k}{(2\pi)^3}\frac{k'^2\d k'}{(2\pi)^3}\sum_{\ell_1m_1}\sigma_{\ell_1m_1}\langle\Phi_{\ell m}^k{\tilde{\delta}_{\ell m}}^{k'*}\rangle G_{\ell_1m_1\ell m}^{\ell m} \nonumber\\
&= C_{\ell m}^{\Phi\tilde{\delta}}\sum_{\ell_1m_1}\sigma_{\ell_1m_1}G_{\ell_1m_1\ell m}^{\ell m} \nonumber\\
&=2\sqrt{15\pi}i^\ell\frac{p}{N_\ell}\left(\int\frac{k^2\d k}{(2\pi)^3}a_{k}^* P_{\hat\delta\hat\delta}(k)J_\ell^W(k)\right)\sum_{\ell_1m_1}\sigma_{\ell_1m_1}G_{\ell_1m_1\ell m}^{\ell m}.
\end{align}
We can simplify this expression by utilizing the full form of $G_{\ell_1m_1\ell m}^{\ell m}$:
\be G_{\ell_1m_1\ell m}^{\ell m} = (-1)^{m_1+1}(2\ell+1)\sqrt{\frac{2\ell_1+1}{4\pi}}\begin{pmatrix}
\ell_1 & \ell & \ell\\m_1 & m & m
\end{pmatrix}\begin{pmatrix}
\ell_1 & \ell & \ell\\2 & 0 & -2
\end{pmatrix}. \ee
This is 0 unless $\rvert \ell_1 \rvert \geq \rvert m_1 \rvert$, $m_1=-2m$, and $0 \leq \ell_1 \leq 2\ell$, and hence,
\be
\label{clmvdelta} 
C_{\ell m}^{V\tilde{\delta}} = 2\sqrt{15\pi}i^\ell\frac{p}{N_\ell}\left(\int\frac{k^2\d k}{(2\pi)^3}a_{k}^* P_{\hat\delta\hat\delta}(k)J_\ell^W(k)\right)\sum_{\ell_1=\lvert 2m \rvert}^{2\ell}\sigma_{\ell_1-2m}G_{\ell_1 -2m \ell m}^{\ell m} .
\ee
This is the 21cm-I--CMB-V cross-correlation for the model of  \cite{Kamionkowski:2018syl}.

\subsection{Towards An Estimator for CMB-V-21cm-I Cross-Correlation}

Our analysis has been purely theoretical, and we have not touched upon a procedure for the extraction of this information from real data.  To do this it would be useful to have a quantity that is easy to measure (i.e., has a low error associated with its measurement), and which is not 0 when $C_{\ell m}^{V\tilde{\delta}}\neq0$. More precisely, we seek to construct a minimum-variance estimator of the CMB-V-21cm-I cross-correlation.

 To do this, we follow a method described in e.g. \cite{Chakraborty:2019sty}. We define the estimator $F_\ell$ to be the weighted average over values of $m$ of the $C_{\ell m}^{V\tilde{\delta}}$,
\be
 F_\ell = \sum_{m=-\ell}^\ell w_{\ell m}C_{\ell m}^{V\tilde{\delta}},
  \ee
where we require $\sum_{m=-\ell}^\ell w_{\ell m}=1$. The minimum-variance estimator is given by an inverse-variance-weighted mean.  This general result can be derived in the present context by computing the set of weights $w_{\ell m}$ that minimize $\delta F_\ell$. To do this, we note that $\delta F_\ell$ is minimized when $ \sum_{m=-\ell}^l w_{\ell m}^2{\delta C_{\ell m}^{V\tilde{\delta}}}^2$ is minimized. Adding a Lagrange multiplier so that our condition will be satisfied, we wish to minimize
\be 
L = \sum_{m=-\ell}^\ell w_{\ell m}^2{\delta C_{\ell m}^{V\tilde{\delta}}}^2 - \lambda\left(\sum_{m=-\ell}^\ell w_{\ell m}-1\right).
 \ee
It can be shown that this is minimized when
\be
 \lambda = 2 \left[ \displaystyle\sum_{m_1=-\ell}^\ell \frac{1}{{\delta C_{\ell m_1}^{V\tilde{\delta}}}^2} \right]^{-1} \;\; , \;\;
 w_{\ell m} = \left[\displaystyle\sum_{m_1=-\ell}^\ell \frac{{\delta C_{\ell m}^{V\tilde{\delta}}}^2}{{\delta C_{\ell m_1}^{V\tilde{\delta}}}^2}\right]^{-1},
 \ee
and hence
\be
\label{flform} 
 F_\ell = \frac{\displaystyle\sum_{m=-\ell}^\ell \frac{C_{\ell m}^{V\tilde{\delta}}}{{\delta C_{\ell m}^{V\tilde{\delta}}}^2}}{\displaystyle\sum_{m=-\ell}^\ell \frac{1}{{\delta C_{\ell m}^{V\tilde{\delta}}}^2}} ,
 \ee
which is precisely an inverse-variance-weighted sum over $m$.

 The task remains to compute $\delta C_{\ell m}^{V\tilde{\delta}}$. The three sources of experimental error for $C_{\ell m}^{V\tilde{\delta}}$ are $a_k$, $P_{\hat\delta\hat\delta}$, and $\sigma_{\ell_1,-2m}$. Assuming all variables are approximately Gaussian, \eqref{clmvdelta} gives
\bea
\displaystyle\left(\frac{\delta C_{\ell m}^{V\tilde{\delta}}}{C_{\ell m}^{V\tilde{\delta}}}\right)^2 = &&
\displaystyle\left(\int\frac{k^2\d k}{(2\pi)^3}a_{k}^* P_{\hat\delta\hat\delta}(k)J_\ell^W(k)\right)^{-2}  \displaystyle\int\d k\left(\frac{k^2}{(2\pi)^3}J_\ell^W(k)\right)^2\left(\delta a_k^{*^2}P_{\hat\delta\hat\delta}(k)^2+a_k^{*^2}\delta P_{\hat\delta\hat\delta}(k)^2\right)  \nonumber \\
 && + \displaystyle \left(\sum_{\ell_1=\rvert 2m \rvert}^{2\ell}\sigma_{\ell_1,-2m} G_{\ell_1,-2m,\ell,m}^{\ell m}\right)^{-2} \displaystyle \sum_{\ell_1=\rvert 2m \rvert}^{2\ell}\left(\delta\sigma_{\ell_1,-2m} G_{\ell_1,-2m,\ell,m}^{\ell m}\right)^2. 
\eea
To find $\delta\sigma_{\ell_1,-2m}$, let us suppose that we calculate $\sigma_{\ell m}$ by dividing the sky into $N$ equal patches and averaging $P_{\ell m}^2$ over these patches, so that $\displaystyle\sigma_{\ell m}=\sqrt{\frac{1}{N}\sum_{\mathrm{patches}}P_{\ell m}^2}$. Then the relative error is,
\be 
\frac{\delta\sigma_{\ell m}}{\sigma_{\ell m}} = \frac{1}{2}\frac{\displaystyle\delta\sum_{\mathrm{patches}}P_{\ell m}^2}{\displaystyle\sum_{\mathrm{patches}}P_{\ell m}^2}  = \frac{\displaystyle\sqrt{\sum_{\mathrm{patches}}\left(P_{\ell m}\delta P_{\ell m}\right)^2}}{\displaystyle\sum_{\mathrm{patches}}P_{\ell m}^2}. 
\ee
Assuming that the RMS $\delta P_{\ell m}$ is constant across all patches, we get
\be
 \delta\sigma_{\ell m} = \sqrt{\frac{\displaystyle\sum_{\mathrm{patches}}\left(P_{\ell m}\delta P_{\ell m}\right)^2}{\displaystyle N\sum_{\mathrm{patches}}P_{\ell m}^2}} = \frac{\delta P_{\ell m}}{\sqrt{N}},
  \ee
where $\delta P_{\ell m}$ on the right-hand side is understood as the RMS value.  In the limit $N\to\infty$, $\delta\sigma_{\ell m}\to0$. Thus if we calculate $\sigma_{\ell m}$ using very many patches, then
\bea
\displaystyle\delta C_{\ell m}^{V\tilde{\delta}} = && 2\sqrt{15\pi}i^\ell\frac{p}{N_\ell}\sqrt{\int\d k\left(\frac{k^2}{(2\pi)^3}J_\ell^W(\chi)\right)^2\left(\delta a_k^{*^2}P_{\hat\delta\hat\delta}(k)^2+a_k^{*^2}\delta P_{\hat\delta\hat\delta}(k)^2\right)} \nonumber \\
&& \;\;\; \cdot \sum_{\ell_1=\rvert 2m \rvert}^{2\ell}\sigma_{\ell_1,-2m} G_{\ell_1,-2m,\ell,m}^{lm}.
 \eea
Plugging this into \eqref{flform}, one finds, 
\be
 F_\ell = \frac{\displaystyle\sum_{m=-\ell}^\ell \frac{C_{\ell m}^{V\tilde{\delta}}}{\displaystyle\left(\sum_{\ell_1=\rvert 2m \rvert}^{2\ell}\sigma_{\ell_1,-2m} G_{\ell_1,-2m,\ell,m}^{\ell m}\right)^2}}{\displaystyle\sum_{m=-\ell}^\ell \frac{1}{\displaystyle\left(\sum_{\ell_1=\rvert 2m \rvert}^{2\ell}\sigma_{\ell_1,-2m} G_{\ell_1,-2m,\ell,m}^{\ell m}\right)^2}}.
 \ee
This constitutes a minimum variance estimator of the cross-correlation $C_{\ell m}^{V\tilde{\delta}}$, as generated by the model  \cite{Kamionkowski:2018syl}. We emphasize this is specific to the form \eqref{clmvdelta}, and hence this is a model-specific estimator, which allows for an estimation of $C_{\ell m}^{V\tilde{\delta}}$ once the individual components in \eqref{clmvdelta} have been estimated directly from data using conventional methods. One could alternatively attempt to construct an estimator to extract $C_{\ell m}^{V\tilde{\delta}}$ itself directly from data; we leave this interesting possibility to future work.

\section{Axions and CMB Cross-Correlation}
\label{sec:axionsVT}

As a final exercise in V-mode physics, here we consider the $TV$ cross-correlation of the CMB, and to do this, we continue with the total angular momentum framework utilized in section \ref{sec:21cm}. We note again that despite being generated by parity-violating processes, the V Stokes parameter of the CMB is a \emph{scalar} \cite{Kosowsky:1994cy}, and the multipole expansion into $Y_{\ell} ^m$ enjoys the same parity properties as that for $T$. This allows for non-vanishing $TV$ cross-correlation. As a concrete example, we consider a model in which axions are the only significant common source of $V$ and $T$, and we shall examine how axions manifest in the $TV$ angular power spectrum.

Let us define $T_{\textrm{axions}}=T-T_{\textrm{w/o}}$ to be the component of the temperature fluctuation that can be attributed to axions. It follows that,
\begin{align}
T_{\textrm{axions},k} &= \left(\Gamma_+-1\right)\left\lvert A'_{+\textrm{w/o}} \right\rvert^2 + \left(\Gamma_--1\right)\left\lvert A'_{-\textrm{w/o}} \right\rvert^2 \nonumber\\
\label{taxions} &= \frac{2\Gamma_+\Gamma_--\Gamma_+-\Gamma_-}{2\Gamma_+\Gamma_-}T_k+\frac{\Gamma_+-\Gamma_-}{2\Gamma_+\Gamma_-}V_k.
\end{align}
The subscripts of $k$ are added as a reminder that this equation refers to individual Fourier modes of the $T$ and $V$ signals, i.e.
\be T=\int\frac{d^3k}{(2\pi)^3}T_ke^{i\vec{k}\cdot\vec{x}}, \ee
and analogously with $V$ and $T_{\textrm{axions}}$. From the plane-wave expansion,
\be 
e^{i\vec{k}\cdot\vec{x}}=\sum_{\ell m}\Psi_{\ell m}^kY_{\ell m}^*(\hat{k}), 
\ee
it follows that $T_{\ell m}^k=T_kY_{\ell m}^*(\hat{k})$, and analogously with $V$ and $T_{\textrm{axions}}$. Thus \eqref{taxions} becomes
\be T_{\textrm{axions,}\ell m}^k=\frac{2\Gamma_+\Gamma_--\Gamma_+-\Gamma_-}{2\Gamma_+\Gamma_-}T_{\ell m}^k+\frac{\Gamma_+-\Gamma_-}{2\Gamma_+\Gamma_-}V_{\ell m}^k. \ee
We have assumed that axion-photon interaction is the only significant common source of $V$ and $T$, which means that the correlation of $V$ with $T$ should be the same as the correlation of $V$ with $V_{\textrm{axions}}$. Thus we get that the $TV$ angular power spectrum is given by,
\begin{align}
\left\langle T_{\ell m}^kV_{\ell m}^{k*}\right\rangle &= \left\langle T_{\textrm{axions,}\ell m}^kV_{\ell m}^{k*}\right\rangle = \frac{2\Gamma_+\Gamma_--\Gamma_+-\Gamma_-}{2\Gamma_+\Gamma_-}\left\langle T_{\ell m}^kV_{\ell m}^{k*} \right\rangle + \frac{\Gamma_+-\Gamma_-}{2\Gamma_+\Gamma_-}\left\langle V_{\ell m}^kV_{\ell m}^{k*} \right\rangle \nonumber\\
&= \frac{\Gamma_+-\Gamma_-}{\Gamma_++\Gamma_-}\left\langle V_{\ell m}^kV_{\ell m}^{k*} \right\rangle \\
\label{TAMresult} C_\ell^{TV} &= \int\frac{k^2\d k}{(2\pi)^3}\frac{\Gamma_+-\Gamma_-}{\Gamma_++\Gamma_-}\left\lvert V_{\ell m}^k\right\rvert^2.
\end{align}
We see that $C_\ell^{TV}$ is given by a quantity that is quite similar to $C_\ell^{VV}$, except for a factor of $(\Gamma_+-\Gamma_-)/(\Gamma_++\Gamma_-)$ inside the integration, which is general $k$-dependent. 

The relation of $C_\ell^{TV}$ and $C_\ell^{VV}$ provides a relation for V-modes generated by axions. This can be expressed as
\be
{\cal P}_{\ell} ^{TV}(k) = \frac{\Gamma_+-\Gamma_-}{\Gamma_++\Gamma_-}  \, {\cal P}_{\ell} ^{VV}(k),
\ee
with ${\cal P}_{\ell}(k)$ the power spectrum of a multipole $\ell$. For the example of adiabatic evolution \cite{Finelli:2008jv}, discussed above \eqref{adiab}, this prefactor is simply $g \phi'/k$. Independent measurements of $ C_{\ell} ^{TV}$ and $ C_{\ell} ^{VV}$ thus uniquely specify the axion coupling and evolution.

\section{Conclusion}
\label{sec:discussion}

In this work we have endeavored to perform a systematic analysis of circular polarization of the cosmic microwave background, and the cross-correlation of this with 21cm cosmology. We have derived the Boltzmann hierarchy for scalar, vector, and tensor modes, of Stokes V parameter, and derived an analytic solution to the Boltzmann hierarchy in the presence of a general source. We then collected proposals in the literature and mapped them onto this description, providing the corresponding source term in the Boltzmann equation.  We put this forward as a roadmap of the science that can be done with future studies of V-modes and their impact on the cosmic microwave background.

In our analysis of 21cm cosmology, we have considered only a single mechanism of CMB-V-21cm-I cross correlation. More generally, one may expect that any $V$ that is generated by propagation of CMB photons across cosmological distances may be correlated with the 21cm intensity, since both populations of photons must traverse the cosmos.  Given the swath of 21cm experiments to be launched in the near future, it is imperative to perform a thorough exploration of model space, namely the set of mechanisms which can generate such a cross-correlation. 

We have also considered $TV$ cross-correlation in the CMB. In contrast with $TB$, this can be non-vanishing even in the absence of additional parity violation. This is due to the simple fact that $V(\theta)$ and $T(\theta)$ are both scalar quantities \cite{Kosowsky:1994cy}. We find axion models lead to a relation between the $TV$ and $VV$ correlations, which relates the spectra and the axion velocity and coupling. This suggests that V may be a useful element in the suite of observational probes of axion.

There are many directions for future work; here we have but scratched the surface. An obvious next step is build a V-mode polarization module to interface with known Boltzmann solvers like CLASS \cite{2011JCAP07034B} or CAMB \cite{Lewis:2002ah}. This will allow for a quantitative analysis of the V-mode polarization. From there one may consider additional fields and interactions, incorporated as additional fluid components to the universe and collision terms in the V-mode Boltzmann equation.   It will also be interesting to perform a full analysis of the vector and tensor modes of V, which we did not develop beyond the Boltzmann hierarchy. 

We also note that many mechanisms to generate V-modes also generate B-modes, e.g.~through the corresponding production of chiral gravitational waves, and vice versa \cite{Inomata:2018rin}. Chiral gravitational waves play an important role in models of 
of leptogenesis  \cite{Alexander:2004us, Maleknejad:2014wsa,Maleknejad:2016dci, Adshead:2017znw, Caldwell:2017chz} and dark matter \cite{Alexander:2018fjp}, and also serve as a complementary signal to the dark matter production in \cite{Maleknejad:2019hdr}. It will be interesting to understand the extent to which V-modes may be a probe of more general baryogenesis and dark matter models. We leave this, and topics discussed above, to future work.

\acknowledgments 
The authors thank Steven J. Clark and Simon Foreman for useful discussions and insightful comments. AP is supported by NASA ROSES NNH17ZDA001N-ATP Grant No. 80NSSC18K10148.

 \newpage
 
\appendix

\section{Solution to the Boltzmann hierarchy with a general source}
\label{app:boltzmann}
To show that \eqref{solutionwithsource} solves \eqref{hierarchywithsource}, one must use the result that
\be \label{besselsumidentity} \sum_{n=0}^{\floor{\frac{\ell'}{2}}} A_n^{\ell'}j_\ell^{(\ell'-2n)}(0) = \delta_{\ell\ell'}, \ee
where $A_n^{\ell'}$ and $\Sigma^{\ell'} _{kn}$ are defined in \eqref{Anldef} and \eqref{sigmaknldef}. In this appendix we present a proof for \eqref{besselsumidentity}.

The spherical Bessel function has a series expansion
\be j_\ell(x)=\sum_{k=0}^\infty\frac{2^\ell(-1)^k(\ell+k)!}{k!(2\ell+2k+1)!}x^{\ell+2k}. \ee
Taking the $\ell'^{\textrm{th}}$ derivative, we find
\be j_\ell^{(\ell')}(x)=\sum_{k=\textrm{max}\left(0,\ceil{\frac{\ell'-\ell}{2}}\right)}^\infty\frac{2^\ell(-1)^k(\ell+k)!(\ell+2k)!}{k!(2\ell+2k+1)!(\ell-\ell'+2k)!}x^{\ell-\ell'+2k}. \ee
When we evaluate at $0$, we will only get the term corresponding to $\ell-\ell'+2k=0$, so
\begin{align} 
\label{besselderivativeidentity}
j_\ell^{(\ell')}(0)&=\left\{\begin{tabular}{c c} 0 & $\ell>\ell'$ or $\ell'-\ell$ odd\\
$\frac{2^\ell(-1)^{\frac{\ell'-\ell}{2}}\left(\frac{\ell'+\ell}{2}\right)!\ell'!}{\left(\frac{\ell'-\ell}{2}\right)!(\ell'+\ell+1)!}$ & $\ell\leq\ell'$ and $\ell'-\ell$ even \end{tabular}\right.\\
&= \sum_{k=0}^{\ceil{\frac{\ell'-\ell}{2}}}\frac{2^\ell(-1)^k(\ell+k)!(\ell+2k)!}{k!(2\ell+2k+1)!}\delta_{\ell,\ell'-2k}
\end{align}
If $\ell>\ell'$ or $\ell'-\ell$ is odd, then \eqref{besselsumidentity} follows trivially from \eqref{besselderivativeidentity}. If $\ell'-\ell$ is even, so that $\frac{\ell'-\ell}{2}\leq\floor{\frac{\ell'}{2}}$, then
\begin{align}
\label{penultimatestep}
\sum_{n=0}^{\floor{\frac{\ell'}{2}}} A_n^{\ell'}j_\ell^{(\ell'-2n)}(0) &= \sum_{n=0}^{\frac{\ell'-\ell}{2}} A_n^{\ell'}j_\ell^{(\ell'-2n)}(0)\nonumber\\
&= \sum_{n=0}^{\frac{\ell'-\ell}{2}} A_n^{\ell'}\sum_{k=0}^{\frac{\ell'-\ell}{2}-n}\frac{2^\ell(-1)^k(\ell+k)!(\ell+2k)!}{k!(2\ell+2k+1)!}\delta_{\ell,\ell'-2(n+k)}\nonumber\\
&= \sum_{n=0}^{\frac{\ell'-\ell}{2}} A_n^{\ell'}\sum_{k=n}^{\frac{\ell'-\ell}{2}}\frac{2^\ell(-1)^{k-n}(\ell+k-n)!(\ell+2k-2n)!}{k!(2\ell+2k-2n+1)!}\delta_{\ell,\ell'-2k}\nonumber\\
&= \sum_{n=0}^{\frac{\ell'-\ell}{2}} A_n^{\ell'}\sum_{k=n}^{\frac{\ell'-\ell}{2}}\frac{2^{\ell'-2k}(-1)^{k-n}(\ell'-k-n)!(\ell'-2n)!}{k!(2\ell'-2k-2n+1)!}\delta_{\ell,\ell'-2k}\nonumber\\
&= \sum_{k=0}^{\frac{\ell'-\ell}{2}} \delta_{\ell,\ell'-2k} \sum_{n=0}^k A_n^{\ell'}\Sigma_{kn}^{\ell'}
\end{align}
From the definition of $A_n^{\ell'}$, it is easy to show that $\displaystyle \sum_{n=0}^k A_n^{\ell'}\Sigma_{kn}^{\ell'}=\delta_{k0}$:
\begin{align}
&\sum_{n=0}^k A_n^{\ell'}\Sigma_{kn}^{\ell'} = A_k^{\ell'}\Sigma_{kk}^{\ell'}+\sum_{n=0}^{k-1} A_n^{\ell'}\Sigma_{kn}^{\ell'} = 0\hspace{1cm}\textrm{if }k>0 \\
&\sum_{n=0}^k A_n^{\ell'}\Sigma_{kn}^{\ell'} = A_0^{\ell'}\Sigma_{00}^{\ell'} =1\hspace{3.15cm}\textrm{if }k=0
\end{align}
Plugging this into \eqref{penultimatestep} yields \eqref{besselsumidentity}.

\section{The effect of axions on $V$}
\label{app:axions}
Following the definitions in section \ref{Definitions}, the $T$ and $V$ Stokes parameters can be expressed in terms of $A_\pm$; specifically,
\be V = \frac{1}{a^4}\left(\rvert A'_+\rvert^2-\rvert A'_-\rvert^2\right) \hspace{2cm} T = \frac{1}{a^4}\left(\rvert A'_+\rvert^2+\rvert A'_-\rvert^2\right) \ee
We can define $V_{\textrm{w/o}}$ and $T_{\textrm{w/o}}$ analogously, except with $A_{\pm\textrm{w/o}}$ instead of $A_\pm$; these are the values $V$ and $T$ would have, if there were no photon-axion interaction. Substituting in \eqref{scaling}, we get
\be\label{Vintermsof} V = \frac{\Gamma_++\Gamma_-}{2}V_\textrm{w/o} + \frac{\Gamma_+-\Gamma_-}{2}T_\textrm{w/o}. \ee
We will assume that $T_{\textrm{w/o}}\approx T$, that is, that axion-photon interaction does not contribute significantly to the temperature fluctuations of the CMB. This is a reasonable assumption, as we know that $C_\ell^{TT}$ is quite well predicted using standard cosmology, without including axions. Also, on a more practical level, if axions \emph{did} significantly affect $T$, then we wouldn't have to use circular polarization to look for evidence of axions in the first place. Taking the derivative, we get
\begin{align}
V' &= \frac{\Gamma'_++\Gamma'_-}{2}V_\textrm{w/o} + \frac{\Gamma'_+-\Gamma'_-}{2}T + \frac{\Gamma_+-\Gamma_-}{2}T' + \frac{\Gamma_++\Gamma_-}{2}V'_\textrm{w/o} \nonumber\\
&= \frac{\Gamma'_++\Gamma'_-}{\Gamma_++\Gamma_-}V + \frac{\Gamma_-\Gamma'_+-\Gamma_+\Gamma'_-}{\Gamma_++\Gamma_-}T + \frac{\Gamma_+-\Gamma_-}{2}T' + \frac{\Gamma_++\Gamma_-}{2}V'_\textrm{w/o},
\end{align}
where in the second line we have substituted for $V{_\textrm{w/o}}$ using \eqref{Vintermsof}.

Note that the assumption that $T_{\textrm{w/o}}\approx T$, i.e. that $T_{\textrm{axions}}$ is negligible, while necessary to derive \eqref{axionsources}, is not used in section \ref{sec:axionsVT}.

\bibliography{VmodeRefs}
\bibliographystyle{JHEP}

\end{document}